\documentclass[12pt]{article}
\usepackage{epsf}
\usepackage{amsmath, amssymb}
\usepackage{graphicx}

\begin{document}

\newcommand{\bi}{\bfseries\itshape}

\renewcommand{\thefootnote}{\fnsymbol{footnote}}
\setcounter{footnote}{0}
\begin{titlepage}

\def\thefootnote{\fnsymbol{footnote}}

\begin{center}

\hfill UT-11-37\\
\hfill IPMU-11-0187\\
\hfill November, 2011\\

\vskip .75in

{\Large \bf 
  Focus Point Assisted by Right-Handed Neutrinos
}

\vskip .75in

{\large
Masaki Asano$^{(a)}$, Takeo Moroi$^{(b,c)}$, Ryosuke Sato$^{(b,c)}$
\\
and Tsutomu T. Yanagida$^{(b,c)}$
}

\vskip 0.25in

{\em $^{(a)}$
II. Institute for Theoretical Physics, University of Hamburg, \\
Luruper Chausse 149, DE-22761 Hamburg, Germany
}

\vskip 0.1in

{\em $^{(b)}$Department of Physics, University of Tokyo,
Tokyo 113-0033, Japan
}

\vskip 0.1in

{\em $^{(c)}$Institute for the Physics and Mathematics of the Universe,\\
University of Tokyo, Kashiwa 277-8568, Japan}

\end{center}
\vskip .5in

\begin{abstract}

  The large hadron collider experiments have now reached the focus
  point region in which the scalar masses are multi-TeV. We study the
  parameter region of the focus point scenario which may realize a
  natural electroweak symmetry breaking avoiding serious fine
  tuning. We show that the region with a mild tuning of $3-5\ \%$
  level is expanded by introducing right-handed neutrinos in the
  framework of the seesaw scenario.  We discuss the prediction of the
  Higgs mass, bounds on the squark and gluino masses, and the relic
  density of the lightest neutralino in such a parameter region.

\end{abstract}

\end{titlepage}

\renewcommand{\thepage}{\arabic{page}}
\setcounter{page}{1}
\renewcommand{\thefootnote}{\#\arabic{footnote}}
\setcounter{footnote}{0}

\section{Introduction}

The low energy supersymmetry (SUSY) is one of the leading candidates
for physics beyond the standard model. In the SUSY models, the
quadratic divergence in the Higgs mass squared term disappears and the
electroweak symmetry breaking scale is arising from SUSY breaking
parameters and SUSY Higgs mass parameter, $\mu$. Thus, it is plausible
that the SUSY particles will be discovered around the electroweak
symmetry breaking scale. However, the LHC experiments recently
reported strong lower bounds on masses of SUSY particles
\cite{Aad:2011qa,Chatrchyan:2011zy}.  For instance, the gluino and
squarks lighter than 1 TeV was already excluded if their masses are
nearly equal.  Such a LHC bound imposes a serious fine tuning problem
to SUSY models.

With the tension between the LHC bound on the masses of superparticles
and fine tuning being given, the focus point scenario
\cite{Feng:1999mn,Feng:1999zg, Feng:1999sw} is one of the interesting
possibilities to consider.  (See also \cite{Feng:2000bp}.)  In the
focus point region, although scalar masses are multi-TeV, the value of
the up-type soft SUSY breaking Higgs mass squared, $m_{H_u}^2$, is
around electroweak scale squared due to the ``focus point'' behavior
of the $m_{H_u}^2$ running; consequently the electroweak symmetry
breaking may be naturally realized by small $m_{H_u}^2$ and $\mu$
parameters. In the multi-TeV scalar mass region, the lower bound on
the gluino mass is relaxed and it is about 600 GeV.  Thus, in focus
point scenario, a mild tuning of parameters may be enough to realize
the electroweak symmetry breaking without conflicting the LHC bounds.

As we see below, however, the Higgs mass bound from the LEP experiment
($m_h\geq 114.4\ {\rm GeV}$ \cite{Nakamura:2010zzi}) also puts serious
constraint on the parameter space.  In particular, a large fraction of
the parameter region with a tuning of a few \% is excluded by the
Higgs mass constraint if we adopt the particle content of the minimal
supersymmetric standard model (MSSM).

One of the plausible extension of the MSSM is to introduce heavy
right-handed neutrinos. Even if the neutrino Yukawa coupling is
$\mathcal{O}(1)$, the small neutrino mass can be explained by the
seesaw mechanism \cite{Yanagida:1979as, GellMann:1976pg,
  Minkowski:1977sc} with the large Majorana mass of the right-handed
neutrino.  The purpose of this letter is to show that if the Yukawa
coupling of a neutrino is $\mathcal{O}(1)$, the allowed parameter
space in the focus point region significantly expands and we have a
large region of mild tuning.  In particular, we show that a parameter
region with $3-5$ \% tuning is allowed in the focus point region where
multi-TeV scalars exist.

\section{Focus point scenario without seesaw}

First, we briefly summarize how the focus point region is constrained
by various experimental bounds in the framework of the MSSM.

In the MSSM, the electroweak symmetry breaking scale is given by the
SUSY Higgs mass parameter, $\mu$, and soft SUSY breaking masses as
\begin{eqnarray}
  \frac{1}{2}m_Z^2 = 
  - \mu^2 + \frac{m_{H_d}^2 - m_{H_u}^2 \tan^2 \beta}{\tan^2 \beta -1}
  \sim - \mu^2 - m_{H_u}^2,
   \label{eq:FT1}
\end{eqnarray}
where $m_{H_u}^2$ ($m_{H_d}^2$) is the up- (down-) type soft SUSY
breaking Higgs mass squared at the electroweak scale and $\tan\beta =
\langle H_u^0 \rangle / \langle H_d^0 \rangle$.  The last relation
holds for a moderately large value of $\tan\beta$ and the relation can
be rewritten by the following form:
\begin{eqnarray}
  \frac{1}{2}m_Z^2 \sim - \mu^2 - m_{H_u}^2 |_{\rm mess} - \delta m_{H_u}^2,
\label{eq:mz}
\end{eqnarray}
where the $m_{H_u}^2 |_{\rm mess}$ is the up-type Higgs mass at the
SUSY breaking mediation scale and $\delta m_{H_u}^2$ denotes the
contribution of the running from the mediation scale to the
electroweak scale.  In naive discussion, naturalness requires that
each term in the right-hand side should not be much larger than the
electroweak symmetry breaking scale.
This requires that the masses of scalar top quarks (stops) should not
be much larger than the electroweak scale in order for the electroweak
symmetry breaking to naturally happen
(for details of the naturalness bound on stop mass, see, for e.g.,
\cite{Chacko:2005ra,Nomura:2005qg,Kitano:2006gv} ).

The above conclusion may change in the focus point region.  To see
this, we study the running of MSSM parameters in the framework of the
minimal supergravity (mSUGRA) with the following input parameters:
\begin{eqnarray}
 \left( m_0, m_{1/2}, A_0, B_0, \mu  \right), \nonumber
\end{eqnarray}
where $m_0$, $m_{1/2}$, $A_0$ and $B_0$ are the universal scalar mass,
gaugino mass, tri-linear coupling and the dimensionful Higgs mixing
parameter at the grand unified theory (GUT) scale $M_{\rm GUT}$,
respectively.  Notice that $B_0$ is determined once the low energy
parameter $\tan\beta$ (as well as other GUT scale parameters) is
fixed.  In mSUGRA, eq.\ (\ref{eq:mz}) becomes
\begin{eqnarray}
  \frac{1}{2}m_Z^2 \sim - \mu^2 - c_{0} m_0^2 
  - c_{1/2} m_{1/2}^2 - c_A A_0^2 - c_{Am} (m_{1/2} A_0),
\end{eqnarray}
where the coefficients $c_{0}$, $c_{1/2}$, $c_A$ and $c_{Am}$ are
determined once the gauge and Yukawa coupling constants, $M_{\rm
  GUT}$, and $\tan\beta$ are given.  The focus point mechanism works
if $c_{0}$ is much smaller than $1$
\cite{Barbieri:1987fn,Chan:1997bi,Feng:1999hg,Feng:1999mn,Feng:1999zg,
  Feng:1999sw}; in such a case, $m_{H_u}^2$ at the electroweak scale
is insensitive to $m_0$ because the focus scale of the $m_{H_u}^2$
running becomes close to the electroweak scale.  This means the
electroweak scale also becomes insensitive to the parameter $m_0$, so
$m_0$ (i.e., typical scalar masses) can be multi-TeV without
conflicting with the fine tuning constraint.  Using the GUT scale and
top quark mass suggested by experimental data, it is well known that
the value of $c_{0}$ is relatively small.  Notice that, on the
contrary, $m_{1/2}$, $A_0$ and $\mu$ should not be much larger than
the electroweak scale for the naturalness.

\begin{figure}[t]
  \begin{center}
    \includegraphics[width=8cm]{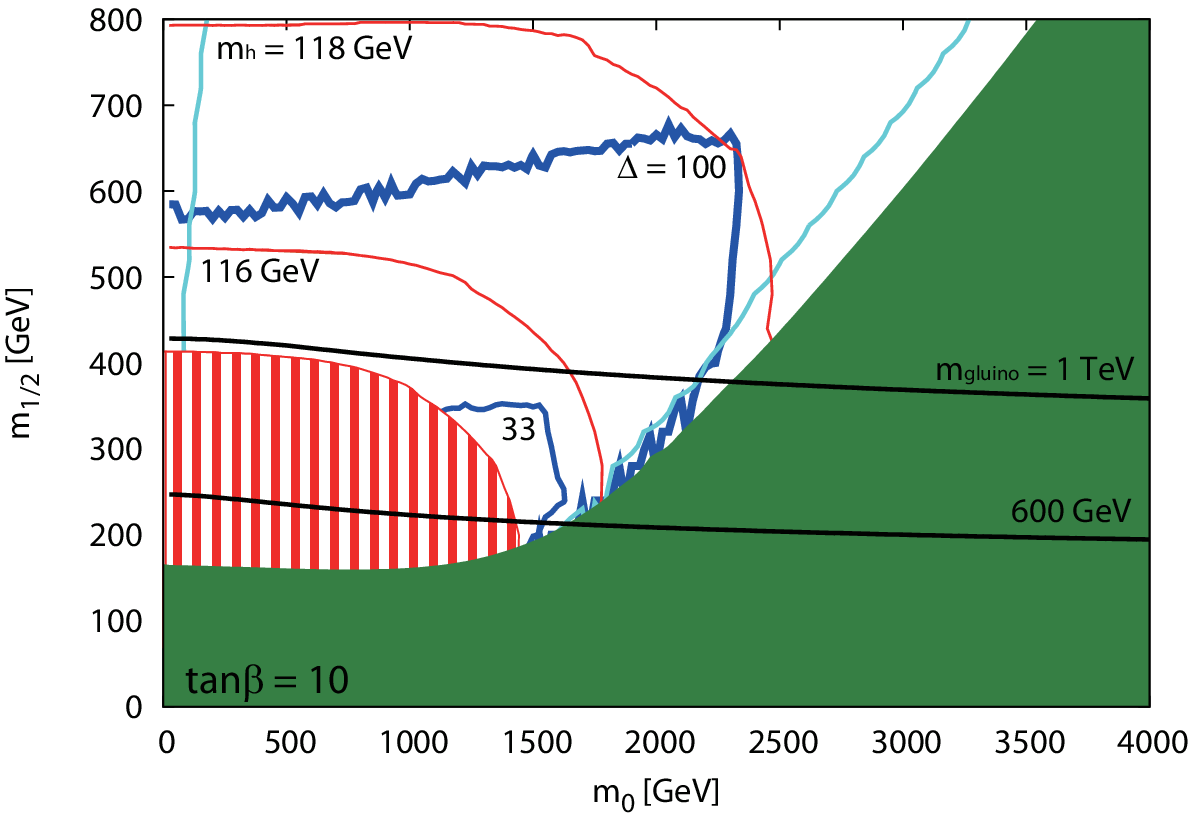}
    \includegraphics[width=8cm]{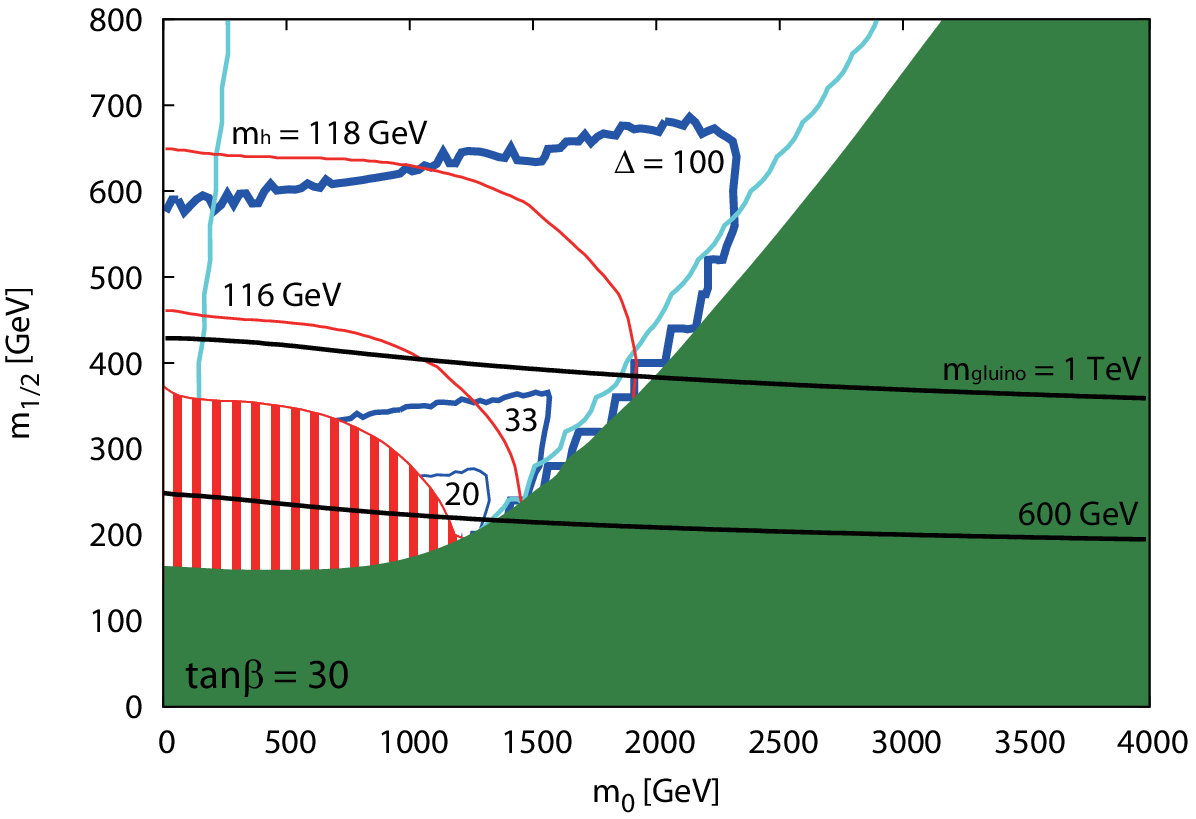}
    \caption{\small Experimental bounds and fine tuning parameter on
      $m_0$ vs. $m_{1/2}$ plane for $\tan\beta =10$ (left) and $30$
      (right) in the framework of the MSSM.  Here, we take $A_0 = 0$.
      The blue lines show contours of fine tuning parameter $\Delta =
      100$, $33$ and $20$ from above.  The green shaded region
      corresponds to the region excluded by the chargino mass limit of
      $103.5$ GeV.  The red stripe regions is the region excluded by
      Higgs mass limit of $114.4$ GeV, while red contours are for the
      Higgs mass of $m_h=116\ {\rm GeV}$ and $118\ {\rm GeV}$ from
      below.  The black line shows contour of $m_{\tilde{g}}= 600$ GeV
      and 1 TeV.  The turquoise line shows the region where the
      thermal relic density of the lightest neutralino becomes
      consistent with the dark matter density suggested by the WMAP
      $\Omega_{c} h^2 = 0.112$ \cite{Komatsu:2010fb}.  }
   \label{fig:FT}
  \end{center}
\end{figure}

In the actual situation, however, the Higgs mass bound imposes a
serious constraint on such a scenario.  This is because a large value
of the stop mass is required to enhance the lightest Higgs mass, which
conflicts with the naturalness bound.  In Fig.\ \ref{fig:FT}, we show
the contours of constant Higgs mass on $m_0$ vs.\ $m_{1/2}$ plane, as
well as the fine tuning parameter defined as
\cite{Ellis:1986yg,Barbieri:1987fn}
\begin{eqnarray}
  \Delta \equiv {\rm max} \left( \Delta_a \right), \qquad
  \Delta_a \equiv \left| \frac{\partial \ln m_Z^2 }{\partial \ln a} \right|,
\end{eqnarray}
with
\begin{eqnarray}
  a = \left( m_0, m_{1/2}, A_0, B_0, \mu \right).
\end{eqnarray}
Notice that $\Delta$ parametrizes the sensitivity of the electroweak
scale to the high scale model parameters.  In the same figure, we also
show contours of constant chargino and gluino masses.  In addition, we
also draw the contour on which the thermal relic abundance of the
lightest neutralino is consistent with the WMAP value $\Omega_c
h^2=0.112$ \cite{Komatsu:2010fb}.  In our analysis, we take
$\tan\beta=10$ and $30$, $M_{\rm GUT}=2\times 10^{16}$ GeV, and the
top quark mass is taken to be $m_t=173\ {\rm GeV}$.  The
renormalization group evolution and SUSY mass spectrum are calculated
by using ISAJET 7.81 \cite{Paige:2003mg}.\footnote
{ In ISAJET code, we modified the function {\tt SSRSGT} which computes
  the threshold correction of the top Yukawa coupling constant $y_t$
  at the SUSY scale.  The original code overestimates this correction,
  which leads to slightly large $y_t$ above the SUSY scale.  It
  affects significantly electroweak symmetry breaking condition in
  large $m_0$ region.  We checked that modified code is almost
  consistent with other codes.  }
In addition, the lightest Higgs mass and the relic abundance are
calculated by FeynHiggs 2.8.5 \cite{Heinemeyer:1998yj} and DarkSUSY
5.0.5 \cite{Gondolo:2004sc}, respectively.

In Fig.\ \ref{fig:FT}, the green shaded region is excluded by the
chargino mass limit from LEP experiments \cite{LEP}.  (Notice that the
green region also includes the parameter region where $m_{H_u}^2$
becomes positive, resulting in the failure of radiative electroweak
symmetry breaking.)  On the other hand, the fine tuning parameter
$\Delta$ is mostly determined by $m_{1/2}$ and $\mu$ when $m_{1/2}$ is
large or by $m_0$ in large $m_0$ region.  Then, even in focus point
region, there exists an upper bound on $m_0$ (and hence on the scalar
masses) once an upper bound on $\Delta$ is imposed.  Such a
naturalness bound contradicts with the Higgs mass bound as seen in
Fig.\ \ref{fig:FT}.  For example, if we take $\Delta\lesssim 20$,
which corresponds to $\sim 5\ \%$ tuning of the parameters for the
electroweak symmetry breaking, the allowed region consistent with the
Higgs mass constraint is found to be quite small even for
$\tan\beta=30$.  In the following, we see how this changes once the
right-handed neutrinos are introduced.

\section{The focus point scenario with seesaw}

One of the plausible extension of the MSSM is to introduce
right-handed neutrinos in order to explain the small neutrino mass by
seesaw mechanism.  Then, the renormalization group running of
$m_{H_u}^2$ is affected by the neutrino Yukawa coupling constants, and
the focus point behavior may change \cite{Kadota:2009fg}.

In the seesaw scenario, the active neutrino mass matrix is given by
\begin{eqnarray}
  [m_{\nu_L}]_{ij} =  \frac{1}{2} \langle H_u^0 \rangle^2
  \sum_{kl} [Y_\nu]_{ik} [Y_\nu]_{jl} 
  [\mathcal{M}_{\nu_R}]^{-1}_{kl},
\end{eqnarray}
where $Y_\nu$ and $\mathcal{M}_{\nu_R}$ are the neutrino Yukawa matrix
and the Majorana mass matrix of right-handed neutrinos, respectively.
(The indices $i$, $j$, $\cdots$ run $1-3$.)  Notice that, using the
Maki-Nakagawa-Sakata (MNS) matrix $V_{\rm MNS}$ \cite{Maki:1962mu},
$m_{\nu_L}$ is expressed as $m_{\nu_L}=V_{\rm MNS}\hat{m}_{\nu_L}
V_{\rm MNS}^T$, where $\hat{m}_{\nu_L}={\rm
  diag}(m_{\nu_1},m_{\nu_2},m_{\nu_3})$ with $m_{\nu_i}$ being masses
of active neutrinos.  As we will discuss later, the rates of the
lepton-flavor violating processes strongly depend on the origin of the
MNS matrix.

It can be easily seen that the focus scale of $m_{H_u}^2$ parameter
changes if the neutrino Yukawa couplings are ${\cal O}(1)$.  This is because
the neutrino Yukawa coupling constant gives negative contribution to
$m_{H_u}^2$ (at lower scale).  Consequently, the upper bound on $m_0$
from the electroweak symmetry breaking condition shifts to large $m_0$
region.  The above statement holds irrespective of the detailed
structure of the Yukawa and Majorana mass matrices of neutrinos.  So,
let us briefly discuss how it happens by adopting universal Majorana
mass matrix (i.e., $\mathcal{M}_{\nu_R}={\rm
  diag}(M_{\nu_R},M_{\nu_R},M_{\nu_R})$) to make the point clear.
(Implication of such a choice to the lepton flavor violation will be
discussed later.)  Using the active neutrino masses suggested from the
neutrino oscillation experiments, the largest eigenvalue of $Y_\nu$
becomes $\mathcal{O}(1)$ if $M_{\nu_R}$ is around $10^{14}$ GeV.
Then, in models with right-handed neutrinos, the renormalization group
equation of $m_{H_u}^2$ contains the following term:
\begin{eqnarray}
  \frac{d m_{H_u}^2}{d \log Q} = 
  \left[ \frac{d m_{H_u}^2}{d \log Q} \right]_{\rm MSSM} + 
  \frac{y_\nu^2}{16\pi^2}
  \left( m_{H_u}^2 + m_{L_3}^2 + m_{N_3}^2 \right)
  \theta(Q-M_{\nu_R})
  + \cdots,
\end{eqnarray}
where $Q$ is the renormalization scale, the first term is the MSSM
contribution, $y_\nu$ is the largest eigenvalue of the neutrino Yukawa
matrix, and $m_{L_3}^2$ and $m_{N_3}^2$ are SUSY breaking mass squared
of the slepton doublet and right-handed sneutrino which couple to
$y_\nu$, respectively.  (In the above equation, we omit the
contribution of tri-linear coupling, which is irrelevant for our
discussion.)  Here, for simplicity, we have assumed that the
eigenvalues of the neutrino Yukawa matrix is hierarchical and that the
contribution of the largest eigenvalue dominates.  If $y_\nu$ is
sizable, $m_{H_u}^2(Q)$ becomes negative at the scale higher than that
in the case of the MSSM.  Then, the focus scale of $m_{H_u}^2$ becomes
higher compared to the MSSM case, and the region with small $\Delta$
extends to the region with larger $m_0$.  In such a region, the Higgs
mass constraint can be satisfied because of large stop masses.  In
addition, the fine tuning parameter $\Delta$ may be suppressed
because, in the parameter region near the chargino mass bound, the
$\mu$-parameter is small.

\begin{figure}[t]
  \begin{center}
    \includegraphics[width=8cm]{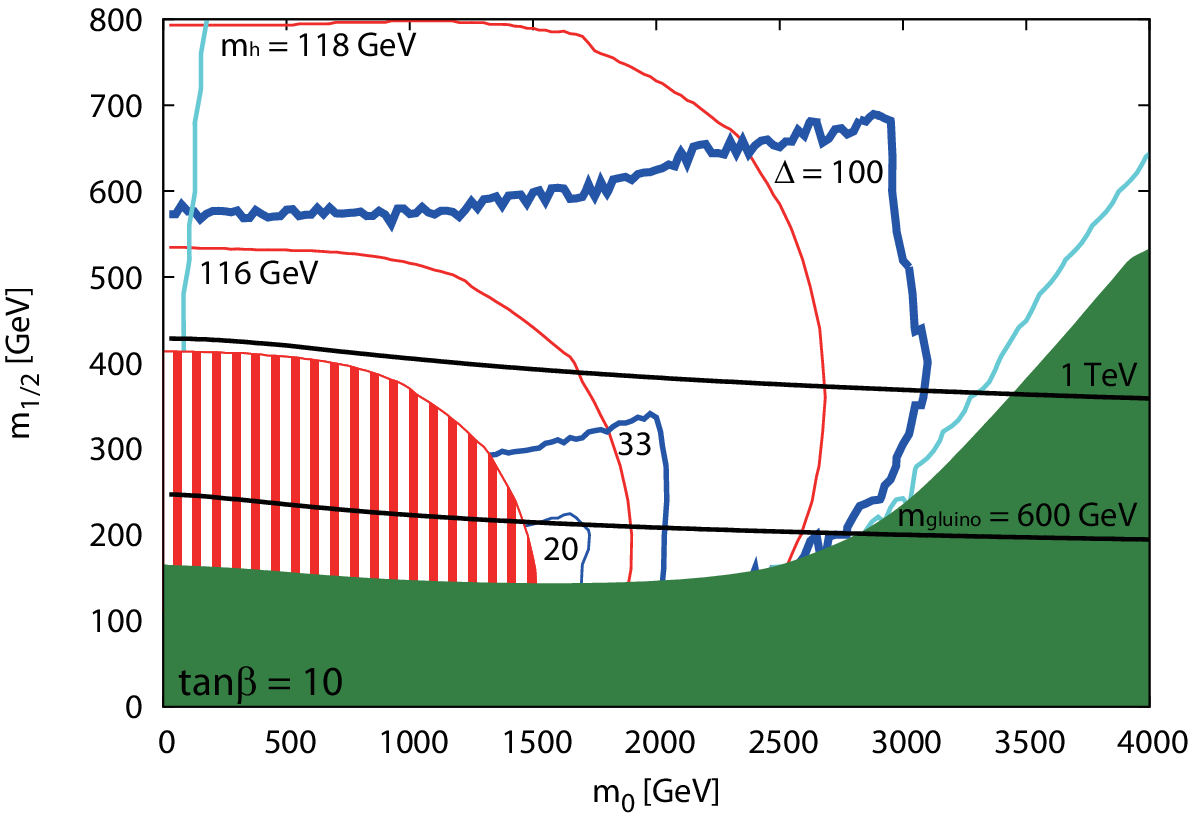}
    \includegraphics[width=8cm]{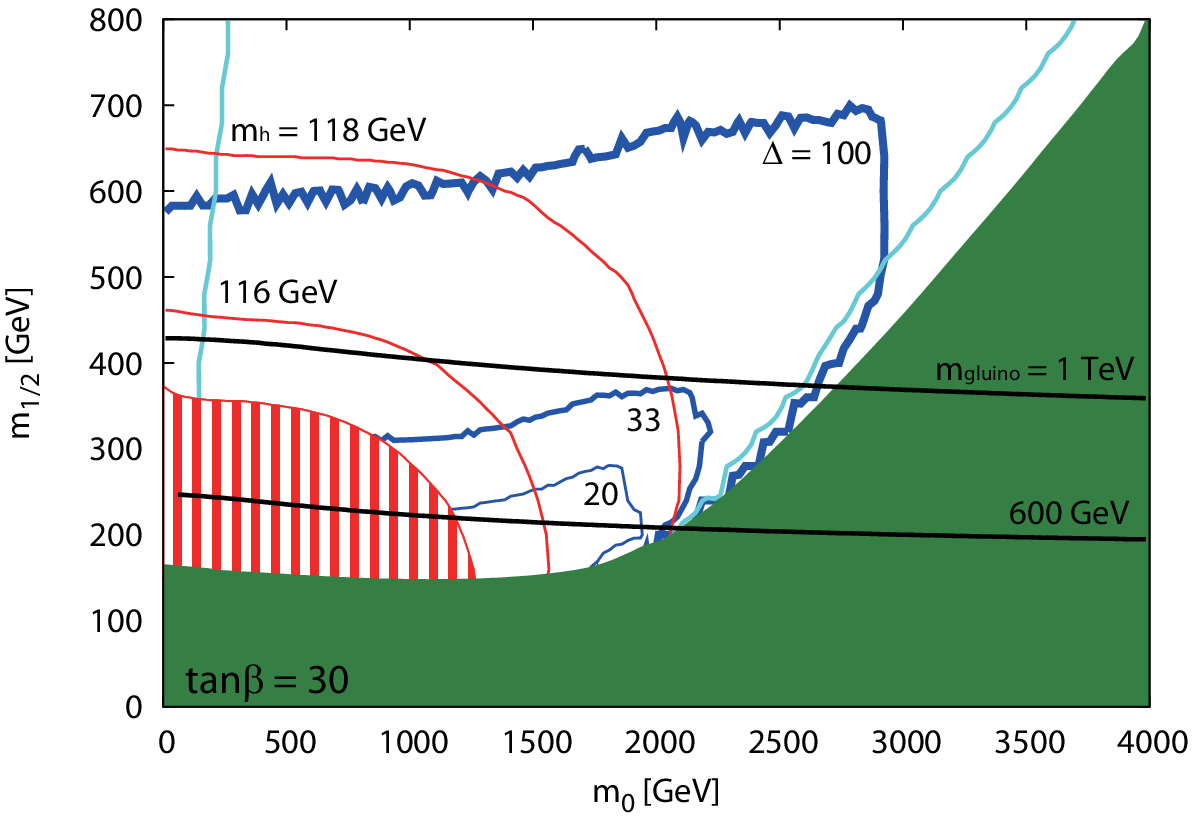}
    \caption{\small Same as Fig.\ \ref{fig:FT}, except for the model
      with right-handed neutrino multiplets.  Here, we take $M_{\nu_R}
      = 2\times 10^{14}$ GeV.}
   \label{fig:FT_RHN}
  \end{center}
\end{figure}

In Fig.\ref{fig:FT_RHN}, we show the contour of the fine tuning
parameter $\Delta$, as well as experimental (and other) bounds.  Here,
we take $A_0 = 0$, $\tan \beta = 10$ and $30$, and $M_{\nu_R} =
2\times 10^{14}$ GeV. The soft SUSY breaking mass squared of the
right-handed sneutrinos are also taken to be $m_0^2$ at the GUT
scale. On the contour of the fine tuning parameter, the horizontal
line is due to $\Delta_{M_{1/2}}$ and $\Delta_{\mu}$ while the
(almost) vertical one is from $\Delta_{m_0}$.  It is remarkable that
the region with a few \% tuning (i.e, $\Delta\lesssim 33$) becomes
larger as we take into account the effect of the neutrino Yukawa
coupling constant.  In the case of $M_{\nu_R} = 2\times 10^{14}$ GeV,
$m_0$ can be as large as $\sim 2\ {\rm TeV}$ even if we require
$\Delta< 33$, which is about $500\ {\rm GeV}$ larger than the case of
the MSSM.  This makes the discovery of the squarks at the LHC
challenging.  Even in such a case, however, it should be noted that a
bound on the $m_{1/2}$ parameter is imposed from the naturalness,
which can be converted to the upper bound on the gluino mass. If we
require $\Delta<33$ (20), gluino should be lighter than 1 TeV (800
GeV) for the case of $\tan\beta=30$.  Thus, for the test of the focus
point scenario, search of the gluino signal is important.  In the
present scenario, it is also notable that the lightest Higgs mass
cannot be so heavy.  Even adopting the tuning of the level of
$\Delta^{-1}=1\ \%$, the Higgs mass is required to be smaller than
about $120\ {\rm GeV}$.

In the present case, one may worry about the thermal relic density of
the lightest neutralino because we consider the parameter region where
the sfermion masses are so heavy that the pair annihilation cross
section of the lightest neutralino is extremely suppressed (if the
lightest neutralino is Bino-like) \cite{Feng:2000gh}.  In
Fig.\ref{fig:FT_RHN}, we also show the contour on which the thermal
relic density of the lightest neutralino becomes consistent with the
present dark matter density.  Notice that, as $m_{1/2}$ becomes
larger, the thermal relic density increases.  In the region that we
are interested in, i.e., in the region where $\Delta^{-1}\lesssim$ a
few \%, the thermal relic density of the lightest neutralino is found
to exceed the present dark matter density.  If we assume the standard
evolution of the universe, this may cause a problem of the
overproduction of the lightest neutralino.  The relic density,
however, depends on the thermal history as well as on the mass
spectrum of SUSY particles.  Thus, we do not take this problem
seriously.  For example, if a sizable amount of entropy is produced
after the freeze-out of the lightest neutralino, this problem can be
avoided.  In addition, if the simple GUT relation among the gaugino
masses does not hold, then the lightest neutralino may not be purely
Bino.  If the lightest neutralino has a sizable Wino or Higgsino
component, the thermal relic density of the lightest neutralino can be
suppressed.  In the scenario of the product-group unification
\cite{Yanagida:1994vq}, such a situation can be easily realized.

Finally, we comment on the lepton flavor violation.  Once we introduce
right-handed neutrinos, it inevitably becomes a new source of lepton
flavor violations.  In particular, even if the slepton mass squared
matrix is universal at the GUT scale, the neutrino Yukawa interaction
induces off-diagonal elements via renormalization group effect
\cite{Borzumati:1986qx}.  (For detailed study of the lepton flavor
violation processes in supersymmetric model with right-handed
neutrinos, see \cite{Hisano:1995cp, Hisano:1998fj}.)  The most
stringent constraint is often from $\mu\rightarrow e\gamma$ process;
the current upper bound on the branching ratio of this process is
${\rm Br} (\mu \to e \gamma) < 2.4 \times 10^{-12}$
\cite{Adam:2011ch}.  Even though the slepton masses are of ${\cal
  O}(1)\ {\rm TeV}$ in the focus point scenario, ${\rm Br} (\mu \to e
\gamma)$ may become large because we consider the case that the
neutrino Yukawa coupling constants are $\mathcal{O}(1)$.  Importantly,
the rates of lepton flavor violating processes strongly depend on the
structures of Yukawa and Majorana mass matrices in the neutrino
sector. For example, if ${\cal M}_\nu$ is universal, the mixing in the
Yukawa matrix should be sizable to realize the mixings of active
neutrinos observed.  Then, it is often the case that ${\rm Br} (\mu
\to e \gamma)$ becomes unacceptably large.  Even in such a case,
however, ${\rm Br} (\mu \to e \gamma)$ depends on the value of
$[V_{\rm MNS}]_{e3}$, which is presently unknown.  If $|[V_{\rm
  MNS}]_{e3}|\ll 1$, ${\rm Br} (\mu \to e \gamma)$ becomes larger than
the experimental bound in the region of $\Delta<100$.  However if
$|[V_{\rm MNS}]_{e3}|\sim 0.06$, there is a possibility of accidental
cancellation so that ${\rm Br} (\mu \to e \gamma)$ is suppressed.
Furthermore, if the neutrino mixing is dominantly from
$\mathcal{M}_{\nu_R}$, the situation may change.  For example, one may
take $[Y_\nu]_{ij}\propto \delta_{ij}$ and
$[\mathcal{M}_{\nu_R}]_{ij}\propto [m_{\nu_L}]^{-1}_{ij}$; then the
flavor violation is significantly suppressed.  This is because the
relevant part of the $\beta$-function of the SUSY breaking mass
squared matrix of slepton is proportional to $[Y_\nu
Y_\nu^\dagger]_{jl}$.  We have checked that, in such a case, the
experimental constraints can be avoided if $m_0\gtrsim 1\ {\rm TeV}$
in the case that the right-handed neutrino masses are around $2\times
10^{14}\ {\rm GeV}$ and $\tan\beta=10$.  More detailed discussion on
this issue will be given elsewhere \cite{AsaMorSatoYan}.

\section{Conclusions and Discussion}

In this letter, we have considered the possibility of relaxing the
fine tuning constraint using the focus point scenario.  We have shown
that the focus point parameter space consistent with the serious Higgs
mass constraint is expanded by introducing right-handed neutrinos.
Due to the contribution from the large Yukawa coupling of the
right-handed neutrinos, the naturalness bound on the $m_0$ vs.\
$m_{1/2}$ plane is changed. Then, we have shown that the parameter
space with a few percent tuning becomes significantly larger.  We have
seen that a parameter space with $\sim 5\ \%$ tuning even exists with
the scalar mass larger than $1\ {\rm TeV}$.  We have found that,
adopting $3 \%$ ($5\ \%$) tuning for the electroweak symmetry
breaking, the gluino mass is bounded above as $m_{\tilde{g}} < 1$ TeV
($800$ GeV).

Finally, we comment on the limit on the gluino mass.  In deriving the
LHC bound on the gluino mass, the GUT relation among the gaugino
masses is usually adopted.  Then, the gluino mass is constrained as
$m_{\tilde{g}} > 600$ GeV in the focus point region. However, in some
class of GUT models, the GUT relation does not always hold; in the
unification model based on product groups \cite{Yanagida:1994vq}, for
example, that is the case \cite{ArkaniHamed:1996jq}. If the masses of
gluino and dark matter are quasi-degenerated, bound on the gluino mass
may be relaxed \cite{gluinolimit}.  Implications of such a possibility
in the focus point scenario (as well as in more general framework)
will be discussed elsewhere \cite{AsaMorSatoYan}.

{\it Acknowledgments:} The authors thank H. Baer for useful
discussion.  This work is supported by Grant-in-Aid for Scientific
research from the Ministry of Education, Science, Sports, and Culture
(MEXT), Japan, No.\ 22244021 (T.M. and T.T.Y.), No.\ 22540263 (T.M.),
and also by the World Premier International Research Center Initiative
(WPI Initiative), MEXT, Japan. The work of R.S. is supported in part
by JSPS Research Fellowships for Young Scientists.  MA acknowledges
support from the German Research Foundation (DFG) through grant BR
3954/1-1.

\end{document}